\documentstyle[emulateapj,psfig]{article}

\newcommand{\be}{\begin{equation}}
\newcommand{\ba}{\begin{eqnarray}}
\newcommand{\ee}{\end{equation}}
\newcommand{\ea}{\end{eqnarray}}  
\newcommand{\etal}{et al.\ }
\def\gtsima{$\; \buildrel > \over \sim \;$}
\def\ltsima{$\; \buildrel < \over \sim \;$}
\def\gsim{\lower.5ex\hbox{\gtsima}}
\def\lsim{\lower.5ex\hbox{\ltsima}}
\def\simgt{\lower.5ex\hbox{\gtsima}}
\def\simlt{\lower.5ex\hbox{\ltsima}}
\def\simpr{\lower.5ex\hbox{\prosima}}

\def\msun{{M_\odot}}

\def\eg{{\frenchspacing\it e.g. }}
\def\spose#1{\hbox to 0pt{#1\hss}}
\def\simpropto{\mathrel{\spose{\lower 3pt\hbox{$\mathchar"218$}}
     \raise 2.0pt\hbox{$\propto$}}}

\begin{document}
\title{AGN Feedback Causes Downsizing}

\author{Evan Scannapieco\altaffilmark{1},  Joseph Silk\altaffilmark{2}, 
Rychard Bouwens\altaffilmark{3}}
\altaffiltext{1}{Kavli Institute for Theoretical Physics,
 Kohn Hall, UC Santa Barbara, Santa Barbara, CA 93106}
\altaffiltext{2}{Astrophysics Department, 
 University of Oxford, Keble Road, Oxford OX1 3RH}
\altaffiltext{3}{Astronomy Department, University of California, 
Santa Cruz, CA 95064}

\begin{abstract}

We study the impact of outflows driven by active galactic nuclei (AGN)
on galaxy formation.  Outflows move into the surrounding intergalactic
medium (IGM) and heat it sufficiently  to prevent it from condensing
onto galaxies.  In the dense, high-redshift IGM, such feedback
requires highly energetic outflows, driven by a large AGN.  However,
in the more tenuous low-redshift IGM, equivalently strong feedback can
be achieved by less energetic winds (and thus smaller galaxies).
Using a simple analytic model, we show that this leads to the {\em
anti-hierarchical} quenching of star-formation in large galaxies,
consistent with current observations.  At redshifts prior to the
formation of large AGN, galaxy formation is hierarchical and follows
the growth of dark-matter halos.  The transition between the two
regimes lies at the $z \approx 2$ peak of AGN activity.

\end{abstract}


\section{Introduction}

Downsizing is not a recent trend.  For the past 10 billion years, the
characteristic stellar mass of forming galaxies has been decreasing,
as demonstrated by a variety of observations.  Space-based
near-ultraviolet (NUV) measurements show that the typical
star-formation rate in galaxies was over an order of magnitude higher
at $z=3$ than at $z=0$ (Arnouts \etal 2005).  Ground-based optical and
near-infrared (NIR) searches indicate that the largest galaxies were
already in place by $z \approx 2,$ while smaller ones continued to
form stars at much lower redshifts (Fontana \etal 2004; Glazebrook
\etal 2004; van Dokkum 2004; Treu \etal 2005).  And both optical and
X-ray surveys detect a steady decrease in the characteristic
luminosity of active galactic nuclei (AGN) below $z \approx 3$ (Pei
1995; Ueda \etal 2003), which is likely to parallel the formation
history of early-type galaxies (\eg Granato \etal 2001).

Yet despite such widespread evidence, galaxy downsizing  (Cowie \etal
1996) was unexpected.  The $\Lambda$ Cold Dark Matter ($\Lambda$CDM)
model, while in spectacular agreement with observations (\eg Spergel
\etal 2003), is a hierarchical theory, in which gravitationally-bound
subunits merge and accrete mass to form  ever-larger objects.
Superimposed on this DM distribution is  the observed baryonic
component.  In canonical models, this gas falls into potential wells,
is shocked heated to the viral temperature, and must radiate this
energy away before forming stars (Rees \& Ostriker 1977; Silk 1977).
The larger the structure, the higher its virial temperature, and the
longer it takes to cool.   Also, in most semi-analytical models, the
dynamical time-scale is adopted for the rate of star formation  via
the Schmidt-Kennicutt law, and this time-scale increases
systematically  with halo mass.  Thus larger galaxies were expected to
form later not only because they occur in later-forming dark matter
halos, but because the cooling and star formation times within such
halos are longer.

Two recent lines of inquiry have  modified this picture.
Theoretically, simulations have shown that cooling is greatly enhanced
by gas inhomogeneities, allowing galaxies to form with stellar masses
over ten times larger than observed (Suginohara \& Ostriker 1998;
Dav\' e \etal 2001).  In fact, virializing shocks may even be
completely absent in smaller halos (Birnboim \& Dekel 2003).
Observationally, it has been discovered that the main heating source
for groups and small galaxy clusters is nongravitational (\eg Arnaud
\& Evrard 1999).  As these are the gaseous halos that were slightly
too massive to form into galaxies, the clear implication is that
nongravitational heating also played a key role  in the history of
large galaxies.

Furthermore, the $\approx 1$ keV per gas particle necessary to preheat
the intracluster medium (ICM) appears to exceed the energy available
from supernovae (\eg Valageas \& Silk 1999; Wu \etal 2000; Kravtsov \&
Yepes 2002).  At the same time, active galactic nuclei (AGN)
are observed to host high-velocity outflows with kinetic luminosities
that may equal a significant fraction of their bolometric luminosity (\eg
Chartas 2002; Pounds \etal 2003; Morganti 2005; see however Sun \etal 2005).
Such outflows would naturally preheat the ICM to 
the necessary levels (Roychowdhury \etal
2004; Lapi \etal 2005), placing this gas at an entropy at which it can
never cool within a Hubble time (Voit \& Brian 2001; Oh \& Benson
2003).  This results in a feedback-regulated picture of galaxy
formation fundamentally different from the canonical approach
(Scannapieco \& Oh 2004; Binney 2004; Di Matteo \etal 2005).

In this letter, we explore the connection between this developing
picture and the widespread observations of downsizing, structuring our
investigation as follows:  In \S2 we develop a simple cooling model
for galaxy formation, which   illustrates how hierarchical formation
naturally arises in the canonical approach.  In \S3 we modify our
model to include AGN feedback, and  show how this leads to a radically
different, anti-hierarchical history.  We compare
these results with recent observations   and conclude with a short
discussion in \S 4.

\section{A Cooling-Regulated Model of Galaxy Formation}

Throughout this study, we assume a $\Lambda$CDM model with parameters $h=0.65$,
$\Omega_0$ = 0.3, $\Omega_\Lambda$ = 0.7, $\Omega_b = 0.05$, $\sigma_8
= 0.87$, and $n=1$, where $h$ is the Hubble constant in units of 100 
km s$^{-1}$ Mpc$^{-1}$,
$\Omega_0$, $\Omega_\Lambda$, and $\Omega_b$
are the total matter, vacuum energy, and baryonic densities in units of the
critical density, $\sigma_8^2$ is the variance of linear fluctuations
on the $8 h^{-1}{\rm Mpc}$ scale, and $n$ is the ``tilt'' of the
primordial power spectrum (\eg Spergel \etal 2003). The Eisenstein \&
Hu (1999) transfer  function is adopted in all cases.

We begin with a model of cooling-regulated galaxy formation
and adopt as simple an approach as possible, to highlight the key 
physical issues.  Unlike more
sophisticated approaches (\eg Kauffmann \etal 1993; 
Somerville \& Primack 1999; Benson \etal 2000), we do not 
attempt to trace the detailed history of a sample of DM halos 
through the use of statistical ``merger trees.''  Instead, we use the fit
described in van den Bosch (2002), which gives the {\em average}
history of a DM halo with a present total mass of $M_0$ as
\be
\log \left[\frac{M(z)}{M_0}\right]= -0.301 
\left[\frac{ {\rm log}(1+z)}{{\rm log}(1+z_{\rm fit})} \right ]^\nu,
\ee
where $\nu \equiv 1.34 + 1.86 \, {\rm log}(1+z_{\rm fit}) - 0.03 \,
{\rm log}(M_0/10^{12} \msun)$, $M(z)$ is the mass of this system at
some redshift $z$, and $z_{\rm fit}$ is defined implicitly through the
relation $D(z_{\rm fit})^{-1} = 1 + 0.40 \sqrt{\sigma^2 (0.254 M_0) -
\sigma^2(M_0)},$ with $D$ being the linear growth factor and
$\sigma^2(M)$ being the variance of linear fluctuations within a
sphere enclosing a total mass $M.$  

Following the canonical picture, we assume that at each redshift
gas flows onto the growing halo along with the dark matter, is shocked to its 
virial density and temperature at that redshift
[$180$ times the mean density and 
$7.2 \times 10^5 \, (M(z)/10^{12} M_\odot)^{2/3} \, (1+z)$ K, respectively], 
and is finally added to the pool of cold gas that forms
stars after a delay of 
\be
t_{\rm cool} = 1.3 \, T_6 \left<n_e \right>^{-1} C^{-1}
\Lambda_{-23}^{-1}(T_6) \qquad {\rm Myrs}.
\label{eq:tcool}
\ee
Here $\left< n_e \right>$ is the average 
number of electrons per cm$^{3}$ in the shocked gas,
$C \equiv \left< n_e^2 \right>  \left<n_e \right>^{-2}$ 
is a ``clumping factor,'' which accounts
for inhomogeneities, $T_6$ is the temperature in units of $10^6$ K,
and $\Lambda_{-23}$ is 
the radiative cooling rate of the gas in units of $10^{-23}$ ergs 
cm$^3$ s$^{-1}.$  
Near 1 keV $ \approx 10^7$ K,
$\Lambda_{-23}$ is roughly a constant
but in general it is a function 
of temperature and metallicity.
For simplicity we adopt the Sutherland \& Dopita (1993) 
equilibrium values for this rate at
a fixed metallicity of $0.1 Z_\odot,$
the level of pre-enrichment of
G dwarf stars in the solar neighborhood (Ostriker \& Thuan 1975)
and in large neighboring galaxies (Thomas \etal 1999).
Finally in keeping with previous semianalytical models, we ignore
gas inhomogeneities, taking $C=1$.  Adopting a larger value,
consistent with simulations, would result in 
the same overall trends, but with galaxies that were much
too large.

\begin{figure}
\centerline{\psfig{figure=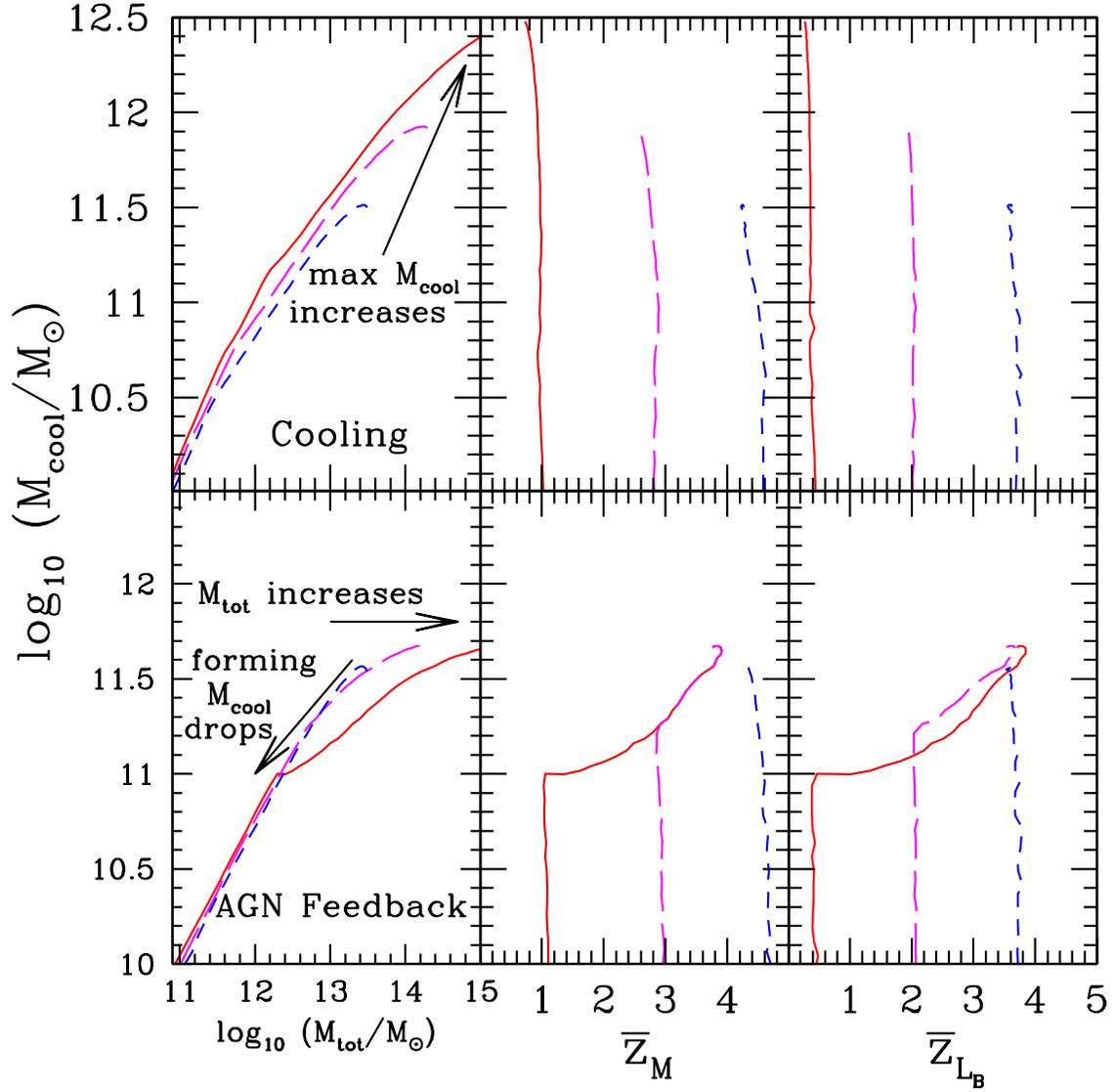,height=16.5cm}}
\caption{Mass of cooled gas in models regulated by cooling (top
panels) and AGN feedback (bottom panels).  In both cases the left
panels compare the cooled gas mass with the total mass, the center
panels give the mass-averaged star-formation redshift, and the right
panels give the luminosity-weighted ($B$-band) average
star-formation redshift.  In all panels the solid, long-dashed, and
short-dashed lines correspond to galaxies observed at redshifts of 0,
1.5, and 3 respectively.  In the cooling-regulated models (\S2),
larger galaxies continue to form at late times, and all galaxies are
dominated by young stars.  In the models regulated by AGN feedback
(\S3), the maximum scale of forming galaxies decreases with time, and
large galaxies are accreted into bigger halos without forming
new stars.  When observed at low redshift, the stars in these objects
are older.}
\label{fig:cooled}
\end{figure}

We then compute the mass of cooled gas as a
function of the total mass and redshift of observation, as illustrated
in Figure \ref{fig:cooled}.  Here we see the classic behavior first
described in Rees \& Ostriker (1977) and Silk (1977).  As their
cooling times are short, $M_{\rm cool}/M_{\rm tot} \approx
\Omega_b/\Omega_0$ in small objects, while for larger masses with
longer cooling times $M_{\rm cool}/M_{\rm tot} \ll \Omega_b/\Omega_0.$
In this picture the maximum $M_{\rm cool}(z)$ is the largest
gas mass that has time to virialize and cool, a value that 
increases with time.  Roughly one can estimate this scaling as
$t_{\rm cool,i}= t_f  \propto (1+z_f)^{-3/2}$  where $t_f$ and $z_f$
are the final time and redshift at which the gas cools while  $t_{\rm
cool,i}$ is computed at the initial redshift at which the gas falls on
the halo.  For the accretion histories described by eq.\ (1),
$T_{\rm vir}$ depends only weakly on redshift, which allows
us to rewrite this as  $M_{\rm cool} \simpropto (1+z_f)^{-3/4}  T_{\rm
vir}.$

In Figure \ref{fig:cooled} we also compute the average redshift of 
formation of 
the resulting stars, adopting both a mass average, that is
${\bar z}_M = z({\bar t}_M)$ where 
\be
{\bar t}_M \equiv \frac{\int_0^{t_{\rm obs}} dt' t' \dot M_{\rm cool}}
{\int_0^{t_{\rm obs}} dt' \dot M_{\rm cool}},
\ee
and a $B$-band luminosity-weighted average, that is
${\bar z}_{L_B} = z({\bar t}_{L_B})$ where 
\be
{\bar t}_{L_B} \equiv 
\frac{\int_0^{t_{\rm obs}} dt' t' \dot M_{\rm cool} 
  \Upsilon_B^{-1}(t_{\rm obs}-t')}
{\int_0^{t_{\rm obs}} dt' \dot M_{\rm cool} 
\Upsilon_B^{-1}(t_{\rm obs}-t')},
\ee
and $\Upsilon_B(t)$ is the $B$-band mass to light ratio of a
population of stars with an age $t,$ assuming a Salpeter initial mass
function as computed by Bruzual \& Charlot (2003).  In all cases
${\bar z}_M$ and ${\bar z}_{L_B}$ are very close to the observed
redshifts, meaning that in objects of all masses and redshifts, most
of the stellar mass and luminosity comes from recently-formed stars.

\section{A Feedback-Regulated Model of Galaxy Formation}

We now modify our model to include AGN
feedback, adopting an approach that parallels 
Scannapieco \& Oh (2004), itself an extension of Wyithe \& Loeb
(2003).  These papers showed that imposing an $M_{\rm BH} \propto
v_c^5$ relationship between black-hole mass and halo circular velocity
(Ferrarese 2002) and assuming that these objects shine at their
Eddington luminosity for a fraction of the dynamical time after each
major merger gives a good fit to the observed AGN luminosity function.  In
particular, each merger with a mass ratio less than 4:1 was associated
with an AGN with a total bolometric energy (in units of $10^{60}$
ergs) of \be
E_{60,{\rm bol}} = 23 M_{12,{\rm cool}}^{5/3} (1+z) \, {\rm ergs}
\label{eq:energy}
\ee
where $M_{12,\rm cool}$ is the mass of cooled gas in units of
$10^{12} \msun.$

Each AGN was then taken to host an outflow, with a 
total kinetic energy equal to $\epsilon_{\rm k} \approx 0.05$ 
of the total bolometric energy.   Adopting a Sedov-Taylor model 
this resulted in a postshock temperature of
$T_{s}(R) = 8.8 \times 10^6 \epsilon_{\rm k} E_{60,\rm bol} 
M^{-1}_{12,{\rm gas}}(R),$
where $M_{12,{\rm gas}}$ is the total gas mass contained with a radius $R$ 
in units of $10^{12} \msun.$  Our simple model here does not distinguish
between merger events, and so instead we relate the {\em change} in 
temperature at a radius $R$ to the {\em change} in cooled mass by 
replacing $M_{12,\rm cool}$ with ${\rm d}M_{12,\rm cool}:$ 
\be
\frac{{\rm d}T_{s}(R)}{{\rm d}M_{12,\rm cool}} =  2.0 \times 10^8 \epsilon_{\rm k} 
\frac{d M_{12,\rm cool}^{2/3} (1+z)} {M_{12,{\rm gas}}(R)} K.
\ee

Apart from this additional contribution to the temperature, our model
exactly parallels that in \S2, with two minor modifications.
First, if a parcel of gas remains at a temperature above the final
virial temperature of the halo for a time longer than its dynamical
time, we assume it escapes from the gravitational potential.  
Secondly, we choose 
the clumping factor to be consistent with simulations, which
show that roughly 40\% of the gas cools onto galaxies by $z=0$ in the
absence of feedback (Dav\'e 2001; Balogh \etal 2001).
This requires us to raise
$M_{\rm cool}$ (without feedback) 
by a factor $\sim$ 30, which corresponds to $C \approx 3.$

With these modifications, and adopting a fiducial value of
$\epsilon_{\rm k} = 0.05$, we constructed plots of  $M_{\rm cool},$
${\bar z}_M$ and ${\bar z_{L_B}}$ as in \S
2.  These are shown in the lower panels of Figure \ref{fig:cooled}.
As in our cooling model, $M_{\rm cool}/M_{\rm tot} \approx
\Omega_b/\Omega_0,$  in small halos, where feedback is weak.  At large
masses, however, feedback causes $M_{\rm cool}$ to evolve  
radically differently.  Now the maximum scale of star-forming galaxies
{\em drops} at lower redshifts.   This is the direct
result of the decrease in the cosmological gas density 
and the fact that at lower redshifts smaller AGNs (and hence smaller
galaxies) can heat this gas to temperatures 
that will not cool within a dynamical time.  The scaling of this
``quenching threshhold'' (Faber \etal 2005) can be estimated by
setting the dynamical time $(\propto n^{-1/2})$ equal to the post-AGN
feedback cooling time given by eqs.\ (\ref{eq:tcool}) \&
(\ref{eq:energy}).  For the $\approx 1$ keV objects we are interested
in this gives $M_{\rm cool} \simpropto (1+z)^{3/4}$.

Galaxies with the largest $M_{\rm cool}$   values ($\approx 3 \times
10^{11} \msun$) are already in place at $z=3$ and continue to fall
into larger halos at lower $z$ without increasing in stellar mass.
Somewhat smaller  $M_{\rm cool} \approx 10^{11} \msun$ galaxies can
recover from an AGN outflow at $z=1.5$, but not at the lowest
redshifts.  Finally galaxies with $M_{\rm cool} < 10^{11} \msun $ have
can hold onto their gas, even at $z=0.$ Note, however, that
our simple models do not include supernova feedback, which is likely
to be important in galaxies with  $M_{\rm tot} \lesssim 10^{11} \msun
$ (Martin 1999).   This mechanism  should behave qualitatively
differently, as it  depends primarily on momentum transfer rather than
heating  (Thacker \etal 2002). Note also that we do not account for
the possibility of initial positive AGN feedback before quenching
occurs, as may be required by observations of efficient star
formation in massive high-redshift galaxies (Silk 2005).

The anti-hierarchical trend in our model 
is also apparent from the mass and $B$-band
luminosity-weighted ages.  In small galaxies, most of the
stars are relatively young and were formed at redshifts close to the
redshifts of observation.  Stars in large galaxies, however, are much
older, and their ages increase as a function of $M_{\rm cool}$.

\section{Comparison with Observations and Discussion}

The features of the simple models developed in \S2 and \S3 can be
directly compared with observations.  As a measure of the typical star
formation rate, we compute the absolute AB magnitude at 1500 \AA \,
for our model galaxies, assuming that $50 \%$ of the cooled gas is in
the form of stars, adopting the Bruzual \& Charlot (2003) population
synthesis models, and ignoring dust.  We do not
attempt to construct a luminosity
function of galaxies, since doing so would require us to use a full
merger-tree formalism.  Rather, for both cooling and AGN feedback
models, we impose a halo mass limit of $\nu(M_{\rm NL},z) = \delta_c
\, \sigma(M)^{-1} D(z)^{-1} \leq 2.5$ at each redshift, to exclude
excessively rare peaks.   For the CDM power spectrum at these
redshifts, $M_{\rm NL} \propto (1+z)^{-6}$, such that the
corresponding virial temperature $T_{\rm vir, NL} \propto (1+z)^{-3}
\propto M_{\rm NL}^{1/2}.$

\begin{table*}
\begin{center}
\caption{\qquad \qquad \qquad \qquad \qquad \qquad \qquad \qquad \qquad
 Properties of Galaxy Formation Models}
\vspace{.05in}
\begin{tabular}{|l|llll|} 
\hline
Model  & Characteristic    & Characteristic  &  Characteristic & $z=0$ Trends \\
       &   $M_{\rm cool}$ at high-$z$ &   $M_{\rm cool}$ at low-$z$    
&  SFR at low-$z$ & \\
\hline
Cooling  &  Tracks  $M_{\rm NL}$      & Cooling-regulated & 
$M_{\rm cool}/t_{\rm dyn}$
 & Bigger galaxies are \\
         &  $\simpropto (1+z)^{-6}$ & $\simpropto (1+z)^{-3/2}$ & 
roughly constant & young and blue \\                                
\hline
AGN  &  Tracks  $M_{\rm NL}$       &  Feedback-regulated & 
$M_{\rm cool}/t_{\rm dyn}$
  & Bigger galaxies are \\
Feedback             &  $\simpropto (1+z)^{-6}$ &  $\simpropto (1+z)^{3/4}$    &  
$\simpropto (1+z)^{9/4}$         & older  and redder\\
\hline
\end{tabular}
\end{center}
\end{table*}

\begin{figure}
\centerline{\psfig{figure=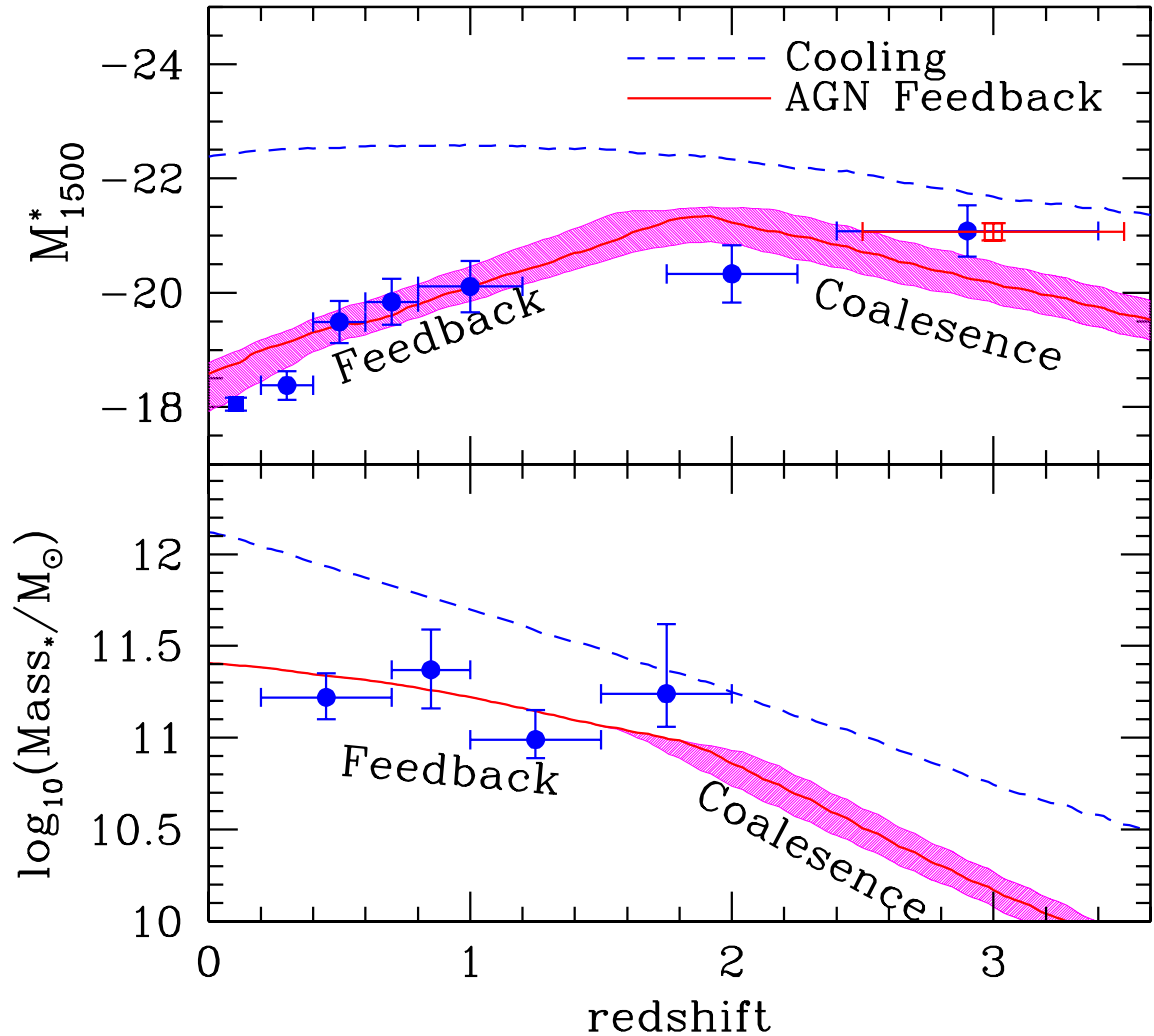,height=14.5cm}}
\caption{{\em Top: } Evolution of $M^*_{1500},$ the characteristic
scale for the instantaneous star formation rate in galaxies.  Lines 
are from our
cooling model (dashed) and $\epsilon_{\rm k}=0.05$ AGN
feedback model (solid) compared with measurements by Arnouts \etal
(2005: solid points) and Steidel \etal (1999: open square).  {\em
Bottom:} Evolution of the characteristic stellar mass as
compared to measurements by Fontana \etal (2004).  Lines are as in the
upper panel.  In both panels, the shaded regions correspond to AGN
feedback models with $\epsilon_{\rm k}$ varied from $0.025$ to $0.1$.
The AGN feedback model is divided into two regimes: a high-redshift
(``coalescence'') regime 
in which the characteristic scale of star-forming galaxies
increases with the nonlinear mass scale, and a low-redshift (``feedback'') 
regime in
which this characteristic scale decreases as a result of quenching from
ever more efficient feedback.}
\label{fig:M*}
\end{figure}

We then find the most luminous galaxy at 1500 \AA \, within the remaining
halos,  and plot its absolute magnitude as a function of redshift in
the upper panel of Figure \ref{fig:M*}.  This is compared
with the observed evolution of $M^*_{1500}$ to $z \sim 3,$ which is
divided into two regions.  At high
redshift, it brightens along with the nonlinear mass scale over the
dynamical time.  At low redshift, the most luminous galaxies no longer lie
in the largest halos.  Rather $M^*_{1500}$ fades along with the
AGN quenching mass scale over $t_{\rm dyn}.$   The redshift at which
these two scales cross marks a distinct transition between
hierarchical and anti-hierarchical growth, which occurs at the peak of
AGN activity (\eg Pei 1995; Ueda \etal 2003).

The shaded regions in this figure show the impact of varying
$\epsilon_{\rm k},$ which shifts the peak value of $M^*_{1500}.$
Modifying our other assumptions ($M_{\rm NL}$,
cold fraction in stars, etc.) would result in similar
changes.  In particular, dust would cause $M^*_{1500}$ to
saturate at $z \approx 2 $ due to the larger attenuation factors
of large starbursts (\eg Wang \& Heckman 1996; Adelberger
\& Steidel 2000; Martin \etal 2005).  Our point is not
to fit these particular parameters to the data, howevever, but rather
that substantial high-redshift brightening and lower-redshift fading of
$M^*_{1500}$ are general and unavoidable features of 
AGN-regulated galaxy formation.

Coincidentally, this sharp peak in $M^*_{1500}$ is now starting to be seen
observationally.  Whereas it has been known for several years that
$M^*$ brightens monotonically from
$z \approx 0$ to $z \approx 2$ in the UV and bluer optical bands 
(\eg Gabasch \etal 2004; Arnouts \etal 2005),
recent measurements at $z \approx 5-6$ (Ouchi \etal 2004;
Bouwens \etal 2005) now also show a downturn
relative to lower redshifts, implying
that $M^*_{1500}$ is most luminous around  $z \approx 3$.
Remarkably, this is
very close to the peak in AGN activity,
suggesting that $z \approx 2-4$ represents the key epoch for gas accretion
onto massive systems.

Turning our attention to the evolution of the maximum stellar mass, as
shown in the bottom panels, similar trends are apparent.  While the
maximum mass in the cooling model increases monotonically, the AGN
feedback model is divided into two regions: one at high redshift in
which the maximum stellar mass increases along with the nonlinear mass
scale,  and one at low redshift in which the maximum stellar mass stays
fixed as the scale at which new galaxies are forming becomes smaller.
The general properties of cooling and AGN feedback models are summarized in
Table 1.

Finally, we point out that not only the temporal, but the {\em
spatial} distribution of AGN in our feedback model is suggestive of
recent observations.  Using over 20,000 galaxies from the 2dF QSO
Redshift Survey, Croom \etal (2005) have studied the spatial
clustering of QSOs near the characteristic scale in the optical
luminosity function.  They measure bias values which, when converted
into $\nu(z)$ values using standard expressions (Mo \& White 1996),
evolve from $2.5 \pm 0.2$ at $z=2.48$ down to $1.1 \pm 0.2$ at
$z=0.56.$ The characteristic scale of low-redshift AGN is downsizing
from the rarest to the most common objects, spreading the heated gas
that extinguishes the formation of galaxies to this day.

\acknowledgements

We thank Biman Nath, Chris Reynolds, and Tomasso Treu for helpful comments.
This work was initiated during a visit by JS to the KITP
as part of the Galaxy-IGM Interactions
Program.  ES was supported by the NSF under
grant PHY99-07949.


\clearpage

\end{document}